%% file: main.tex
\documentclass[journal,12pt,final]{IEEEtran}

\usepackage[utf8]{inputenc}
\usepackage[dvipsnames]{xcolor}
\usepackage{soul}
\usepackage{amsmath}
\usepackage{booktabs}
\usepackage{graphicx}
\usepackage{pgfplots}
\usepackage{pgfplotstable}
\usepgfplotslibrary{fillbetween}
\usepgfplotslibrary{groupplots}
\pgfplotsset{compat=1.17}
\usetikzlibrary{arrows.meta}
\usepackage{subcaption}
\usepackage[font=footnotesize]{caption}
\usepackage{paralist}
\usepackage{amssymb}
\usepackage[range-phrase=--,per-mode=symbol-or-fraction,range-units=single,list-units=single,detect-all]{siunitx}
\usepackage{orcidlink}

\usepackage{cite} 
\usepackage{csquotes}

\usepackage{hyperref}
\usepackage[capitalise,noabbrev]{cleveref}
\Crefformat{figure}{#2Fig.~#1#3}
\Crefformat{equation}{#2Eq.~(#1)#3}

\ifCLASSINFOpdf
\else
\fi

\usepackage{acro}
\usepackage{mfirstuc}
\DeclareAcronym{CFD}{short=CFD, long=computational fluid dynamics}
\DeclareAcronym{EM}{short=EM, long=electromagnetic}
\DeclareAcronym{GNSS}{short=GNSS, long=global navigation satellite systems}
\DeclareAcronym{GPS}{short=GPS, long=global positioning system}
\DeclareAcronym{ISAC}{short=ISAC, long=integrated sensing and communication}
\DeclareAcronym{ISI}{short=ISI, long=inter-symbol interference}
\DeclareAcronym{LiDAR}{short=LiDAR, long=light detection and ranging}
\DeclareAcronym{MC}{short=MC, long=molecular communication}
\DeclareAcronym{RF}{short=RF, long=radio frequency}
\DeclareAcronym{MIMO}{short=MIMO, long=multiple-input multiple-output}
\DeclareAcronym{ML}{short=ML, long=machine learning}
\DeclareAcronym{MOS}{short=MOS, long=metal-oxide semiconductor}
\DeclareAcronym{OFDM}{short=OFDM, long=orthogonal frequency division multiplexing}
\DeclareAcronym{OOK}{short=OOK, long=on-off keying}
\DeclareAcronym{Rx}{short=Rx, long=receiver}
\DeclareAcronym{SLAM}{short=SLAM, long=simultaneous localization and mapping}
\DeclareAcronym{Tx}{short=Tx, long=transmitter}
\DeclareAcronym{TinyML}{short=TinyML, long=tiny machine learning}
\DeclareAcronym{VOC}{short=VOC, long=volatile organic compound}
\DeclareAcronym{5G}{short=5G, long=fifth generation}
\DeclareAcronym{6G}{short=6G, long=sixth generation}

\definecolor{mdblue}{HTML}{396AB1}
\definecolor{mdorange}{HTML}{DA7C30}
\definecolor{mdgreen}{HTML}{3E9651}
\definecolor{mdred}{HTML}{CC2529}
\definecolor{mdgray}{HTML}{535154}
\definecolor{mdpurple}{HTML}{6B4C9A}
\definecolor{mdbrown}{HTML}{922428}
\definecolor{mdyellow}{HTML}{948B3D}
\pgfplotscreateplotcyclelist{colormd}{%
mdblue,mdorange,mdgreen,mdred,mdgray,mdpurple,mdbrown,mdyellow}

\makeatletter
\pgfdeclareplotmark{invtriangle}{
    \pgfpathmoveto{\pgfqpoint{-2pt}{2pt}}
    \pgfpathlineto{\pgfqpoint{2pt}{2pt}}
    \pgfpathlineto{\pgfqpoint{0pt}{-2pt}}
    \pgfpathclose
    \pgfusepathqfillstroke
}
\makeatother

\usepackage{titlesec}

\IEEEoverridecommandlockouts 

\input{definitions}

\begin{document}

\setlength{\skip\footins}{10pt}

\title{VaporISAC: Integrated Sensing and Communication via Molecular Signals}
\author{%
    Sunasheer~Bhattacharjee\,\orcidlink{0000-0002-9878-7328}, 
    Mart\'in~Schottlender\,\orcidlink{0009-0008-7628-2354}, 
    Pit~Hofmann\,\orcidlink{0000-0001-5933-6017},~\IEEEmembership{Member,~IEEE}, 
    Juan~A.~Cabrera\,\orcidlink{0000-0002-7525-2670}, 
    Frank~H.\,P.~Fitzek\,\orcidlink{0000-0001-8469-9573},~\IEEEmembership{Fellow,~IEEE},
    and~Falko~Dressler\,\orcidlink{0000-0002-1989-1750},~\IEEEmembership{Fellow,~IEEE}%
\thanks{%
    S. Bhattacharjee and F. Dressler are with the School of Electrical Engineering and Computer Science, TU Berlin, Germany. E-mail: \{bhattacharjee,dressler\}@ccs-labs.org.
}%
\thanks{%
    M. Schottlender, P. Hofmann, J. A. Cabrera, and F. H. P. Fitzek are with the Deutsche Telekom Chair of Communication Networks and the Centre for Tactile Internet with Human-in-the-Loop (CeTI), TUD Dresden University of Technology, Germany. E-mail: \{martin.schottlender,pit.hofmann,juan.cabrera,frank.fitzek\}@tu-dresden.de.
}%
\thanks{This work was supported  by the German Federal Ministry of Research, Technology and Space (BMFTR) under grant IoBNT 16KIS1986K, the German Research Foundation (DFG) under grant NaBoCom DR 639/21-2,
DFG as part of Germany's Excellence Strategy -- EXC 2050/2 -- Cluster of Excellence ``Centre for Tactile Internet with Human-in-the-Loop'' (CeTI) of TU Dresden under grant number 390696704, and BMFTR as part of the research program Communication Systems ``Souverän. Digital. Vernetzt.'' Joint project 6G-life, grant number 16KIS2413K, and the program “Verbundprojekt: Disruptive Kommunikationsparadigmen für technologische Souveränität, Resilienz und Shared Prosperity - Translation in Industrie und Aufbau innovativer Technologiedemonstratoren - CommUnity,” grant number 16KISS012K.
}%
}

\IEEEaftertitletext{\vspace{-1.5\baselineskip}}

\maketitle
\begin{abstract}
Conventional \ac{EM}-based \ac{ISAC} systems degrade in cluttered, obstructed, and radio-frequency-hostile environments, while macroscopic \ac{MC} remains largely unexplored as an \ac{ISAC} medium.
This article introduces VaporISAC, a molecular \ac{ISAC} framework in which chemical vapor pulses simultaneously convey information and probe the propagation environment, enabling a \emph{one signal, two outputs} paradigm.
The same received waveform is jointly processed to recover transmitted information and infer environmental properties such as airflow, turbulence, smoke, and chemical conditions.
Rather than replacing conventional \ac{EM}-based \ac{ISAC}, VaporISAC complements existing approaches in chemically dynamic, infrastructure-limited, and \ac{EM}-challenged environments.
The sensing principles, system architecture, proof-of-concept demonstrations, emerging applications, and open research challenges of VaporISAC are presented, positioning it as a promising new paradigm for resilient communication and environmental sensing.
\end{abstract}

\IEEEpeerreviewmaketitle

\acresetall

\section{Introduction} 
\label{sec:introduction}

\begin{figure}
    \centering
    \def\svgwidth{\columnwidth}
    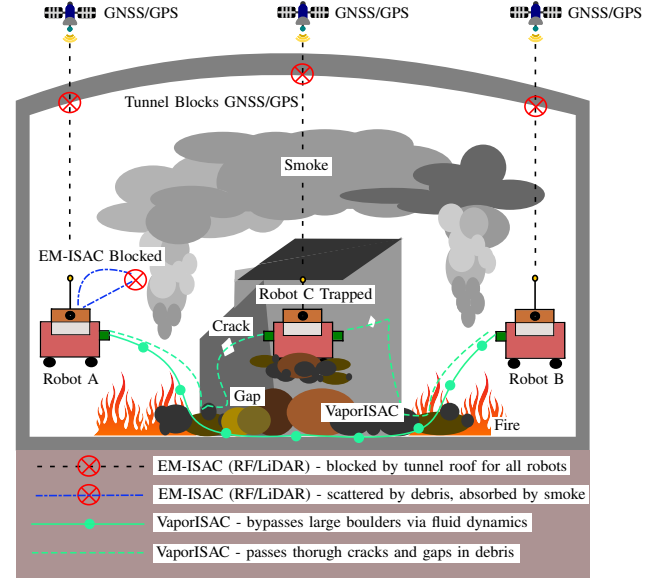
    \caption{Disaster scenario with communication layers.}  
    \label{fig:vaporisac_scenario}
\end{figure}

\Ac{ISAC} is widely recognized as a key enabler of \ac{6G} networks, integrating communication and sensing through shared waveforms and hardware resources.
By jointly enabling data transmission and environmental perception, \ac{ISAC} improves spectral efficiency and supports context-aware applications such as autonomous systems, smart infrastructure, and industrial automation~\cite{nabati2025opportunities}.
However, existing \ac{ISAC} frameworks rely on \ac{EM} signaling, limiting their effectiveness in environments where signal propagation is severely impaired.

Disaster and industrial environments, such as collapsed tunnels, burning buildings, underground pipelines, and smoke-filled facilities, present significant challenges for conventional sensing and communication.
\Ac{RF} signals suffer from severe attenuation and multipath effects, optical sensors degrade in smoke and particulate matter, and \ac{GNSS}/\ac{GPS} is unavailable in enclosed spaces~\cite{yeom2025microwave,bystrov2021experimental,zeeshan2023review}.
These limitations motivate the exploration of alternative physical modalities for \ac{ISAC}, cf.\ Fig.~\ref{fig:vaporisac_scenario}.

In this article, we introduce VaporISAC (cf.\ Fig.~\ref{fig:vaporisac_scenario}), a framework that brings the \ac{ISAC} paradigm to the chemical domain by using controlled vapor releases as both information carriers and environmental probes.
Instead of replacing \ac{RF} systems, VaporISAC complements existing approaches by enabling communication and situational awareness in places where \ac{EM}-based methods become unreliable.
To our knowledge, this is the first framework to unify macroscale \ac{MC} with \ac{ISAC}, establishing a new direction for unified communication and sensing in extreme environments.

\section{EM-ISAC Capabilities and Limits} \label{sec:em-isac}

\Ac{EM}-\ac{ISAC} integrates communication and sensing within a common \ac{RF} platform by sharing spectrum, hardware, and signal processing resources.
Using wireless waveforms, such as orthogonal frequency division multiplexing, \ac{EM}-\ac{ISAC} enables simultaneous data transmission and environmental perception, including object detection, localization, motion estimation, and channel characterization.
Recent advances in sub-6 GHz, mmWave, and massive \ac{MIMO} systems have demonstrated high-rate communication with accurate sensing capabilities~\cite{liu2024millimeter-wave}, establishing \ac{EM}-\ac{ISAC} as a key technology for \ac{6G} applications such as autonomous systems, smart infrastructure, and industrial automation~\cite{zhang2026integrated}.

However, the effectiveness of \ac{EM}-\ac{ISAC} decreases in extreme environments where the propagation medium is highly dynamic or obstructed.
Concrete, debris, and metal introduce severe attenuation, scattering, and multipath effects, degrading communication reliability and sensing accuracy~\cite{yeom2025microwave}.
Although lower-frequency \ac{RF} signals provide improved penetration, they offer limited sensing resolution, whereas mmWave systems are more susceptible to blockage and environmental variations~\cite{batra2025demonstration}.
Similarly, optical modalities such as cameras and \ac{LiDAR} suffer from degradation under smoke, dust, and aerosols, while thermal imaging provides only limited robustness~\cite{bystrov2021experimental,li2024research}.
Furthermore, infrastructure failures can disrupt \ac{GNSS}/\ac{GPS}, cellular, and Wi-Fi connectivity, requiring autonomous platforms to rely on local sensing and navigation mechanisms such as \ac{SLAM}~\cite{zeeshan2023review}.

Beyond these propagation limitations, \ac{EM}-\ac{ISAC} is inherently constrained by its sensing modality.
Radio signals primarily capture \ac{EM} interactions with objects, enabling estimation of parameters such as range, angle, and velocity, but cannot directly measure physical quantities such as airflow, gas concentration, smoke density, or turbulence.
These limitations motivate the exploration of VaporISAC in extreme environments.

\section{Sensing Principles of MC}

\Ac{MC} uses molecules or particles as information carriers instead of \ac{EM} waves.
In air-based systems, signaling molecules are transported through advection and diffusion, causing the received waveform to depend on both the transmitted symbols and environmental factors such as airflow, turbulence, smoke, and obstacles~\cite{atthanayake2018experimental}.
Therefore, molecular signals inherently contain information about both the transmitted message and the surrounding environment.

Despite this dual dependency, most \ac{MC} research has focused primarily on communication, where environmental effects are typically modeled as channel impairments that degrade information transfer~\cite{akyildiz2008nanonetworks,okcu2025smell-preprint}.
However, waveform characteristics such as attenuation, delay, dispersion, and concentration variations also provide valuable information about the propagation environment.

VaporISAC leverages this property by treating molecular waveforms as combined communication and sensing signals.
The received signal can be used not only for data recovery, but also to infer environmental properties, including airflow, turbulence, obstacle-induced flow variations, and aerosol or smoke conditions.
Thus, communication and sensing are extracted from the same molecular propagation process instead of requiring separate sensing modalities.

\begin{table*}
\centering
\caption{Capabilities of electromagnetic- and chemical vapor-based integrated sensing and communication.}
\label{tab:rf_vs_mc_isac}
\begin{tabular}{lll}
\toprule
\textbf{Feature} & 
\textbf{\ac{EM}-\ac{ISAC} (\ac{RF}/mmWave/\ac{LiDAR})} & 
\textbf{VaporISAC (Chemical Vapor)} \\
\midrule

\textbf{Propagation} & 
EM wave propagation & 
Molecular diffusion and/or advection \\
\midrule

\textbf{Communication Range} & 
Long range (\qtyrange{10}{500}{\meter}) & 
Short range (\qtyrange{0.5}{5}{\meter}) \\
\midrule

\textbf{Data Rate} & 
High (Mbps--Gbps) & 
Low (bps) \\
\midrule

\textbf{Smoke and Debris} & 
Strong attenuation and multipath & 
Naturally robust \\
\midrule

\textbf{Penetration} & 
Blocked by sub-wavelength gaps & 
Diffuses freely through cracks and gaps \\
\midrule

\textbf{Environmental Sensing} & 
Geometry, motion, range, velocity & 
Airflow velocity, turbulence, smoke density, chemical hazards \\
\midrule

\textbf{Infrastructure} & 
Base stations/GNSS/GPS & 
Fully ad hoc \\
\midrule

\textbf{Power Consumption} & 
High (W) & 
Low (mW--W) \\
\midrule

\textbf{Best Use Case} & 
Wide-area communication & 
Front-line disaster zones \\
\bottomrule

\end{tabular}
\end{table*}

This concept is supported by recent experimental work that demonstrates alcohol-based \ac{MC} for simultaneous binary decoding and localization of odor-sources using a mobile robot~\cite{yao2025integrated}.
The robot navigates toward an odor source while decoding messages encoded using alcohol molecules and an \ac{OOK} scheme.
However, the sensing capability in this work is mainly limited to source localization and does not explicitly estimate environmental parameters.

Beyond passive environmental observation, VaporISAC can also support reactive sensing, where emitted probe molecules interact with ambient chemical species and generate measurable secondary signatures.
These reactions could enable inference of chemical concentration, spatial distribution, and reactivity, extending sensing beyond physical transport phenomena to chemical state estimation.
In this way, vapor waveforms become active probes of the environment, transforming \ac{MC} from a communication-centric paradigm into a unified framework for communication, environmental sensing, and chemical inference.
A comprehensive comparison between the capabilities of \ac{EM}-\ac{ISAC} and VaporISAC is listed in Table~\ref{tab:rf_vs_mc_isac}.

\section{VaporISAC System Architecture}\label{sec:arch}

\begin{figure}
    \centering
    \def\svgwidth{\columnwidth}
    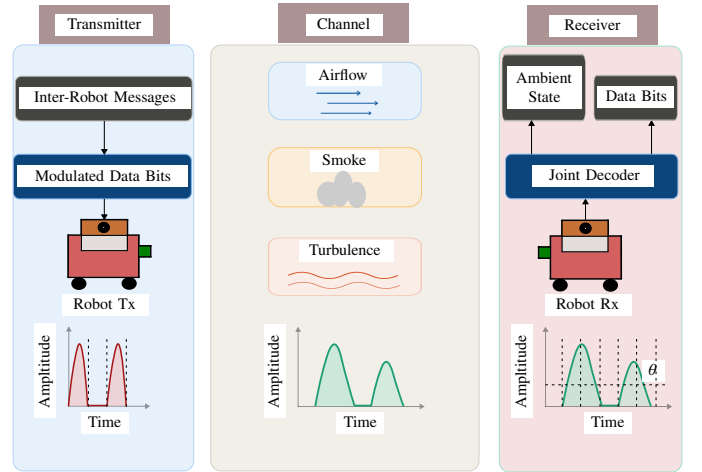
    \caption{Proof-of-concept idea of chemical vapor-based integrated sensing and communication with the transmitter, the channel, and the receiver side components, methods, and influences.}  
    \label{fig:vaporisac_mc_link}
\end{figure}

In the following, we present the proposed VaporISAC architecture, consisting of a chemical vapor \ac{Tx}, a diffusion--advection propagation medium, and a combined communication--sensing \ac{Rx}.
An exemplary proof-of-concept scenario is illustrated in Fig.~\ref{fig:vaporisac_mc_link}.

\subsection{Transmitter}

\Ac{Tx} consists of a low-power microcontroller, a controllable vapor release mechanism (e.g., a micro-pump or spray actuator), and a reservoir containing chemically safe tracer agents.
The microcontroller generates symbol sequences and controls vapor emission timing and quantity using \ac{MC} modulation schemes such as \ac{OOK}, concentration shift keying, molecular shift keying, or their hybrids~\cite{bhattacharjee2022digital}.
Each information symbol is realized as a controlled vapor burst characterized by its release time and emitted mass.
In this sense, \ac{Tx} is analogous to a baseband waveform generator in conventional \ac{ISAC} systems, except that the waveform is physically realized through chemical release rather than \ac{EM} radiation.

\subsection{Propagation Medium}

Vapor propagation is governed by molecular diffusion and bulk airflow, which transform emitted pulses into time-varying concentration waveforms.
Airflow affects arrival time, diffusion and turbulence introduce temporal spreading, and attenuation modifies signal amplitude.
These propagation effects create physically meaningful waveform variations that enable environmental inference.

Unlike conventional sensing systems, VaporISAC does not explicitly require dedicated sensing pilots or preambles, since each transmitted vapor pulse simultaneously conveys information and probes the environment.
This creates an inherent communication-sensing tradeoff: higher symbol rates improve throughput but reduce sensing resolution, whereas longer symbol intervals enhance environmental estimation at the cost of communication rate.

\subsection{Receiver}

\Ac{Rx} employs a chemical sensing front-end, such as a metal-oxide semiconductor sensor or an eNose, to measure time-varying vapor concentrations and convert them into electrical signals containing communication symbols and environmental signatures.
The received waveform is influenced by vapor transport as well as sensor characteristics, including response time, memory, and saturation.

Using a shared sensing and processing pipeline, \ac{Rx} is designed to jointly decode the transmitted information and to estimate effective environmental parameters such as airflow, dispersion, turbulence, and attenuation.
Physics-aware propagation models, statistical methods, or lightweight \ac{ML} approaches can be used to interpret waveform variations as structured environmental information rather than noise.
The proposed framework focuses on low-dimensional environmental inference instead of full spatial reconstruction, enabling efficient operation on resource-constrained platforms and infrastructure-free deployments.

\section{Proof-of-Concept Demonstrations}
\label{sec:poc}

Based on the proposed architecture, this section demonstrates the feasibility of VaporISAC through complementary analytical and experimental results.

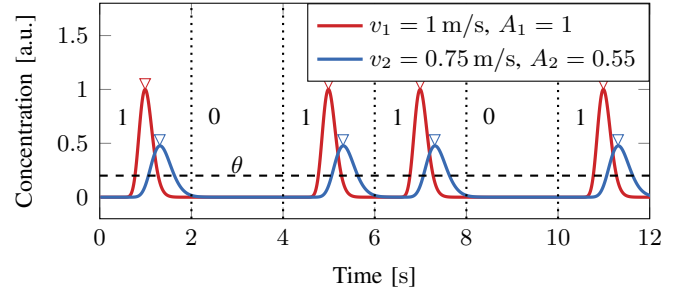
\begin{figure}
    \centering
    {
        \begin{tikzpicture}

            \begin{axis}[
                width=1\columnwidth,
                height=0.5\columnwidth,
                grid=both,
                xmajorgrids=false,
                ymajorgrids=false,
                xmin=0, xmax=12,
                ymin=-0.2, ymax=1.8,
                xlabel={Time [$\si{\second}$]},
                ylabel={Concentration [a.u.]},
                font=\footnotesize,
                legend cell align=left,
                legend style={
                    at={(1,1)},
                    anchor=north east,
                }
            ]

            \addplot [very thick, mdred]
                table[x=Time_s, y=Rx1_Normalized] {plots/isac.dat};

            \addplot [very thick, mdblue]
                table[x=Time_s, y=Rx2_Normalized] {plots/isac.dat};

            \addplot [black, dashed, thick, domain=0:42] {0.2};
            \node[] at (axis cs:3,0.3) {$\theta$};

            \addplot[
                only marks,
                mark=invtriangle,
                mark size=3pt,
                mark options={fill=white,draw=mdred},
                mdred
            ]
            coordinates {
                (0.99,1.05) (4.99,1.05) (6.99,1.05) (10.99,1.05)
            };

            \addplot[
                only marks,
                mark=invtriangle,
                mark size=3pt,
                mark options={fill=white,draw=mdblue},
                mdblue
            ]
            coordinates {
                (1.31,0.53) (5.31,0.53) (7.31,0.53) (11.31,0.53)
            };

            \foreach \x in {2,4,6,8,10} {
                \addplot [black, dotted, thick]
                    coordinates {(\x,-0.2) (\x,1.8)};
            }

            \node[] at (axis cs:0.5,0.75)  {1};
            \node[] at (axis cs:2.5,0.75)  {0};
            \node[] at (axis cs:4.5,0.75) {1};
            \node[] at (axis cs:6.5,0.75) {1};
            \node[] at (axis cs:8.5,0.75) {0};
            \node[] at (axis cs:10.5,0.75) {1};

            \legend{{$v_1 = \SI{1}{\meter\per\second}$, $A_1 = 1$}, {$v_2 = \SI{0.75}{\meter\per\second}$, $A_2 = 0.55$}}

            \end{axis}
        \end{tikzpicture}
    }
    \vspace{-1.5em}
    \caption{On-off keying modulated analytical molecular waveform exploited for both data transmission and environmental sensing.}
    \label{fig:isac_trans}
\end{figure}

\begin{figure}[t]
    \centering
    \includegraphics[width=1.0\linewidth]{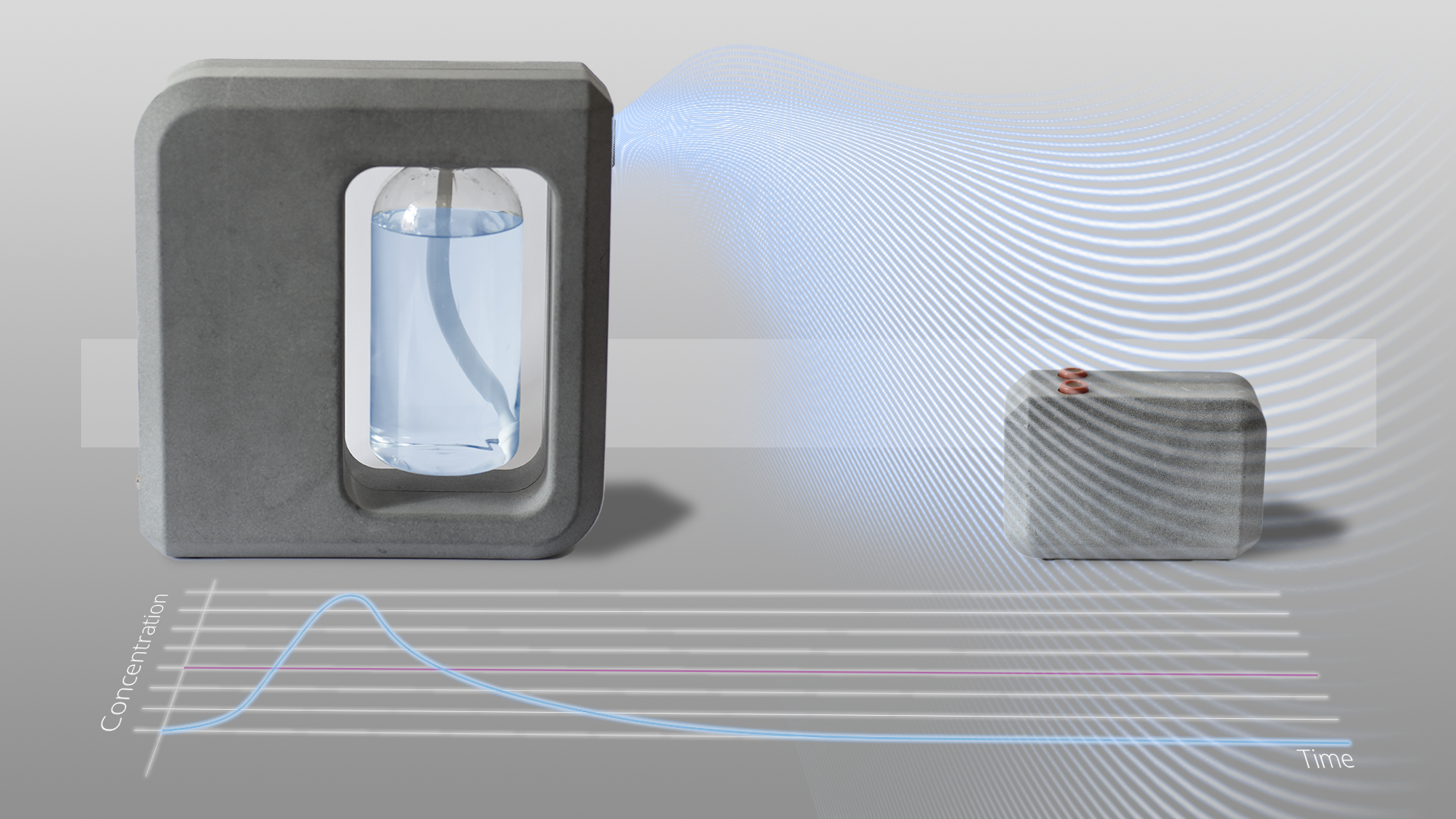}  
    \caption{Experimental setup of the macroscopic molecular communication testbed used for chemical vapor-based integrated sensing and communication. Extracted from \cite{hofmann2022testbed}.}  
    \label{fig:mc_experimental_testbed}
\end{figure}

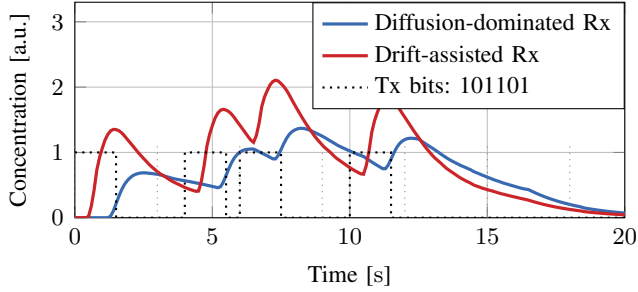
\begin{figure}[t]
    \centering
    \begin{tikzpicture}
    \begin{axis}[
        width=1\linewidth,
        height=0.5\linewidth,
        grid=both,
        xmin=0, xmax=20,
        ymin=0, ymax=3.3,
        xlabel={Time [$\si{\second}$]},
        ylabel={Concentration [a.u.]},
        font=\footnotesize,
        legend cell align=left,
        legend style={
            at={(1,1)},
            anchor=north east,
            fill=white,
            draw=black
        }
    ]

    \addplot [very thick, mdblue]
        table[col sep=space, x=time, y=rx_diffusion_norm]
        {plots/mc_bitstream_101101.dat};
    \addlegendentry{Diffusion-dominated Rx}

    \addplot [very thick, mdred]
        table[col sep=space, x=time, y=rx_drift_norm]
        {plots/mc_bitstream_101101.dat};
    \addlegendentry{Drift-assisted Rx}

    \addplot [thick, black, dotted, const plot]
        table[
            col sep=space,
            x=time,
            y expr=\thisrow{tx_bit}*1
        ]
        {plots/mc_bitstream_101101.dat};
    \addlegendentry{Tx bits: 101101}

    \foreach \x in {0,3,6,9,12,15,18} {
        \addplot [gray, dotted, forget plot]
            coordinates {(\x,0) (\x,1.1)};
    }

    \end{axis}
    \end{tikzpicture}

    \caption{Received concentration response for the transmitted bitstream under diffusion-dominated and drift-assisted propagation.}
    \label{fig:mc_bitstream_rx}
\end{figure}

Fig.~\ref{fig:isac_trans} shows the dual-purpose operation of VaporISAC, where a single molecular waveform supports both communication and environmental sensing.
The concentration profiles are generated using the one-dimensional diffusion--advection channel model~\cite{farsad2016comprehensive} for the transmitted \ac{OOK} sequence \enquote{101101}, assuming a symbol duration of $T_\mathrm{s}=\SI{2}{\second}$ and a \ac{Tx}--\ac{Rx} distance of $x=\SI{1}{\meter}$.
The red and blue waveforms correspond to airflow velocities of $v_1=\SI{1}{\meter\per\second}$ and $v_2=\SI{0.75}{\meter\per\second}$, respectively, with a common diffusion coefficient of $D=\SI{0.01}{\meter^2\per\second}$.
The blue waveform additionally includes attenuation ($A_2=0.55$) to emulate propagation losses caused by environmental obstructions.

Despite the different propagation conditions, a fixed detection threshold ($\theta$) can demonstrably recover the transmitted \ac{OOK} symbols in both cases.
At the same time, peak arrival times (marked by inverted triangles) reveal airflow velocity, while peak amplitudes indicate propagation losses.
Changes in waveform spreading and \ac{ISI} can further provide information about medium dispersion and turbulence.
These results indicate that communication and environmental sensing can, in principle, be performed simultaneously using the same received waveform.

To experimentally validate this concept, we consider the macroscopic \ac{MC} testbed reported in~\cite{hofmann2022testbed}, shown in Fig.~\ref{fig:mc_experimental_testbed}.
The setup employs a spray-based ethanol \ac{Tx} and an MQ-3 chemical sensor operating with \ac{OOK} modulation.
Fig.~\ref{fig:mc_bitstream_rx} shows the normalized received concentration for the same transmitted sequence (\enquote{101101}) under diffusion-assisted and drift-dominated propagation.

Compared with diffusion-assisted propagation, airflow produces concentration peaks that arrive approximately $44\%$ earlier and are $66\%$ stronger.
Using the measured peak shift together with the known \ac{Tx}--\ac{Rx} separation of one meter, an effective drift velocity of approximately \SI{0.314}{\meter\per\second} can be estimated.
These results suggest that environmental parameters can be inferred directly from communication waveforms without dedicated sensing transmissions.

\section{VaporISAC Use Cases}

To illustrate the relevance of VaporISAC, we present applications in indoor robotics, industrial systems, emergency response, and chemically active environments.

\textbf{\ac{GNSS}/\ac{GPS}-Denied Indoor Robotics:} A primary application of VaporISAC lies in indoor environments where \ac{GNSS}/\ac{GPS} is unavailable and radio-based localization suffers from multipath fading, blockage, or attenuation, including basements, warehouses, parking structures, tunnels, and multi-room buildings.
In such settings, robots can exchange navigation commands, localization updates, and map information through molecular signaling while simultaneously exploiting distortions in vapor propagation to infer directional airflow and coarse spatial structure.
These propagation signatures could effectively enable vapor-assisted \ac{SLAM} or \enquote{sniffing out the path}, where the same vapor waveform supports both inter-robot communication and environmental awareness.
This capability is particularly valuable for multi-robot coordination in environments lacking reliable positioning infrastructure.

\textbf{Airflow Monitoring:} Industrial facilities such as chemical plants, refineries, storage units, and manufacturing sites, as well as modern smart buildings, contain complex ventilation systems and enclosed airflow networks that are often difficult to monitor in real time.
VaporISAC could enable autonomous agents to exchange inspection status, leak alerts, airflow maps, occupancy information, and control updates while simultaneously sensing environmental conditions through the same molecular channel.
Variations in vapor propagation delay, attenuation, and dispersion can potentially reveal anomalies such as duct blockages, leakage points, irregular airflow patterns, stagnant air regions, and poorly ventilated zones.
This dual communication--sensing capability could effectively enable coordinated inspection, environmental diagnostics, and airflow characterization without relying on dense sensor deployments or communication infrastructure, supporting applications ranging from industrial monitoring to energy-efficient building management and indoor air quality optimization.

\textbf{Search and Rescue Operations:} Search-and-rescue operations in confined or partially collapsed environments, including disaster sites, tunnels, mining accidents, and fire-affected buildings, are often challenged by obstructed visibility, disrupted communication infrastructure, and irregular airflow.
VaporISAC could enable autonomous agents to maintain communication links for sharing survivor locations, hazard alerts, and navigation instructions while simultaneously using vapor propagation distortions to infer environmental structure and hazard conditions.
Airflow pathways derived from vapor dynamics can also indicate open voids, navigable spaces, smoke-laden regions, and potentially dangerous areas during fire incidents, supporting the localization of trapped individuals, identification of survivable zones, and safer navigation.
Unlike optical or \ac{EM} systems, VaporISAC is expected to remain more robust in heavily obstructed and smoke-filled environments.

\textbf{Chemical Environments and Swarms:} In chemically dynamic environments such as agricultural fields, laboratories, bio-reactive zones, and industrial facilities, VaporISAC can be used to exchange coordination messages and sense environmental conditions through a shared molecular channel.
Vapor waveforms used for inter-agent communication can also be designed as reactive probes that interact with ambient chemical species, producing secondary signatures that could potentially reveal chemical composition, transport dynamics, and potential leakage sources. 
The communication--sensing framework and reactive inference capability can enhance environmental awareness, source localization, and coordination efficiency, enabling more adaptive and resilient swarm behavior.

\section{Open Research Challenges}

We finally discuss key open research challenges in realizing practical VaporISAC systems, spanning modeling, algorithms, hardware, and system integration.

\textbf{Communication--Sensing Tradeoff:} VaporISAC inherently couples sensing and communication, creating a tradeoff between transmission performance and environmental estimation accuracy.
Higher data rates require more frequent emissions and shorter symbol intervals, increasing interference and reducing sensing fidelity.
Conversely, accurate sensing often requires sparse transmissions and longer observation windows, limiting throughput.
Characterizing this tradeoff and developing adaptive strategies that balance sensing and communication under dynamic conditions remain important open challenges.

\textbf{Realistic Channel Modeling:} A major challenge is developing realistic yet tractable models for three-dimensional turbulent molecular propagation.
Simplified diffusion--advection models fail to capture buoyancy, obstacles, thermal gradients, and resulting non-Gaussian plume behavior in disaster environments.
Although \ac{CFD} tools provide high accuracy, their cost prevents real-time use.
Surrogate models trained on \ac{CFD} data could enable real-time inference from sparse measurements on resource-limited platforms.

\textbf{Chemical Multiple Access:} A critical scalability challenge in VaporISAC is the lack of a multiple access framework for multi-agent \ac{MC}. 
In disaster scenarios, simultaneous emissions create overlapping plumes with spatial-temporal interference.
Unlike \ac{RF} systems, \ac{MC} channels lack frequency orthogonality, with interference worsened by nonlinearities, \ac{ISI}, and adsorption-desorption dynamics.
This requires chemical multiple-access strategies, including time scheduling, spatial separation, orthogonal species, and code division multiple access-inspired coding, which are validated in realistic turbulent flows.

\textbf{Hybrid Architecture:} A key challenge in VaporISAC is designing hybrid \ac{EM}-optical-\ac{MC} systems where modalities operate cooperatively rather than independently.
While \ac{EM} sensing primarily provides information such as range, velocity, and obstacle geometry, VaporISAC is designed to infer environmental properties including airflow, smoke, and chemical hazards from the received molecular waveform.
The core open problem is how to dynamically allocate sensing, communication, and decision-making across modalities.
This motivates adaptive fusion frameworks that select or combine channels based on environmental conditions and mission needs, enabling robust operation and improved situational awareness in degraded disaster environments.

\textbf{Decision-Directed Tracking:} VaporISAC relies on decision-directed channel tracking, where parameters are updated from detected emissions for shared communication and environmental estimation.
Its convergence and stability in \ac{MC} channels remain largely unexplored.
Unlike \ac{RF} equalization, it involves stochastic plume evolution, strong non-stationarity, and asymmetric updates.
Decision errors cause cascading effects, e.g., missed detections skip updates, while false alarms corrupt estimates via noise fitting.
Key challenges include stability under airflow dynamics, rapid changes, and uncertainty, as well as soft-decision probabilistic updates.

\textbf{Communication Limits, Coding, and Reliability:} Although current macroscopic \ac{MC} systems achieve only modest data rates suitable for low-rate inter-agent coordination, improving communication reliability and spectral efficiency remains a key research challenge for VaporISAC.
This requires rethinking diffusion-based modulation and coding, including burst-based schemes, molecular multi-carrier signaling, spatial \ac{MIMO} arrays, and error-correcting codes that account for correlated molecular noise, while operating under stringent ultra-low-power constraints.

\textbf{Hardware, Safety, and Validation:} A major barrier to VaporISAC maturation is the lack of experimental validation.
Deployment requires addressing emission variability, sensor drift, nonlinear sensor response, and strong temperature dependence under extreme conditions.
Additionally, mobile robotic integration imposes strict size, weight, and power constraints for emission, sensing, and communication modules.
Safety limits volatile organic compounds below flammability thresholds, motivating non-flammable, low-toxicity alternatives.
A controlled testbed with turbulence, thermal gradients, and smoke is needed to validate models, benchmark performance, and generate reproducible datasets.

\section{Conclusion} \label{sec:conclusion}

This article introduced VaporISAC, a new combined sensing and communication paradigm that extends \ac{ISAC} beyond the \ac{EM} domain by exploiting chemical vapor propagation.
Unlike conventional \ac{EM}-based approaches, VaporISAC leverages the propagation dynamics of molecular signals to simultaneously support reliable data exchange and environmental sensing, making it particularly attractive for cluttered, smoke-filled, infrastructure-denied, and chemically dynamic environments.

We presented the sensing principles underlying \ac{MC}, proposed a collective communication--sensing system architecture, and demonstrated through analytical and experimental proof-of-concept demonstrations that a single molecular waveform can simultaneously convey digital information and reveal environmental properties such as airflow and propagation losses.
Potential applications in autonomous robotics, industrial monitoring, search-and-rescue operations, and chemically active environments further illustrate the versatility of the proposed framework.

Although many challenges remain, including realistic channel modeling, scalable multi-agent communication, hybrid EM--molecular architectures, and experimental validation, VaporISAC establishes a new research direction at the intersection of \ac{MC} and \ac{ISAC}.
We envision future autonomous systems employing multiple complementary physical modalities, where \ac{EM} and molecular signaling operate cooperatively to provide resilient communication and richer environmental awareness in situations where either modality alone would be insufficient.




\bibliographystyle{IEEEtran}
\bibliography{IEEEabrv,references}

\end{document}

%% file: definitions.tex
\newcommand\hide[1]{}

\newcommand{\bitem}{\begin{itemize}}
\newcommand{\eitem}{\end{itemize}}
\newcommand{\benum}{\begin{enumerate}}
\newcommand{\eenum}{\end{enumerate}}
\newcommand{\bcitem}{\begin{compactitem}}
\newcommand{\ecitem}{\end{compactitem}}
\newcommand{\bcenum}{\begin{compactenum}}
\newcommand{\ecenum}{\end{compactenum}}

\newcommand{\x}{$\times$}

%% file: figures/isac_1.pdf_tex
\begingroup%
  \makeatletter%
  \providecommand\color[2][]{%
    \errmessage{(Inkscape) Color is used for the text in Inkscape, but the package 'color.sty' is not loaded}%
    \renewcommand\color[2][]{}%
  }%
  \providecommand\transparent[1]{%
    \errmessage{(Inkscape) Transparency is used (non-zero) for the text in Inkscape, but the package 'transparent.sty' is not loaded}%
    \renewcommand\transparent[1]{}%
  }%
  \providecommand\rotatebox[2]{#2}%
  \newcommand*\fsize{\dimexpr\f@size pt\relax}%
  \newcommand*\lineheight[1]{\tiny\fontsize{\fsize}{#1\fsize}\selectfont}%
  \ifx\svgwidth\undefined%
    \setlength{\unitlength}{595.27559055bp}%
    \ifx\svgscale\undefined%
      \relax%
    \else%
      \setlength{\unitlength}{\unitlength * \real{\svgscale}}%
    \fi%
  \else%
    \setlength{\unitlength}{\svgwidth}%
  \fi%
  \global\let\svgwidth\undefined%
  \global\let\svgscale\undefined%
  \makeatother%
  \begin{picture}(1,0.9047619)%
    \lineheight{1}%
    \setlength\tabcolsep{0pt}%
    \put(0,0){\includegraphics[width=\unitlength,page=1]{figures/isac_1.pdf}}%
    \put(0.19805385,0.49592717){\color[rgb]{0,0,0}\makebox(0,0)[t]{\lineheight{1.29999995}\smash{\begin{tabular}[t]{c}EM-ISAC Blocked\end{tabular}}}}%
    \put(0,0){\includegraphics[width=\unitlength,page=2]{figures/isac_1.pdf}}%
    \put(0.50574232,0.63037887){\color[rgb]{0,0,0}\makebox(0,0)[t]{\lineheight{1.29999995}\smash{\begin{tabular}[t]{c}Smoke\end{tabular}}}}%
    \put(0,0){\includegraphics[width=\unitlength,page=3]{figures/isac_1.pdf}}%
    \put(0.36444441,0.72575395){\color[rgb]{0,0,0}\makebox(0,0)[t]{\lineheight{1.29999995}\smash{\begin{tabular}[t]{c}Tunnel Blocks GNSS/GPS\end{tabular}}}}%
    \put(0,0){\includegraphics[width=\unitlength,page=4]{figures/isac_1.pdf}}%
    \put(0.41771825,0.28585395){\color[rgb]{0,0,0}\makebox(0,0)[t]{\lineheight{1.29999995}\smash{\begin{tabular}[t]{c}Gap\end{tabular}}}}%
    \put(0,0){\includegraphics[width=\unitlength,page=5]{figures/isac_1.pdf}}%
    \put(0.39370894,0.39061206){\color[rgb]{0,0,0}\makebox(0,0)[t]{\lineheight{1.29999995}\smash{\begin{tabular}[t]{c}Crack\end{tabular}}}}%
    \put(0,0){\includegraphics[width=\unitlength,page=6]{figures/isac_1.pdf}}%
    \put(0.51947141,0.43823493){\color[rgb]{0,0,0}\makebox(0,0)[t]{\lineheight{1.29999995}\smash{\begin{tabular}[t]{c}Robot C Trapped\end{tabular}}}}%
    \put(0,0){\includegraphics[width=\unitlength,page=7]{figures/isac_1.pdf}}%
    \put(0.15399697,0.31193124){\color[rgb]{0,0,0}\makebox(0,0)[t]{\lineheight{1.29999995}\smash{\begin{tabular}[t]{c}Robot A\end{tabular}}}}%
    \put(0,0){\includegraphics[width=\unitlength,page=8]{figures/isac_1.pdf}}%
    \put(0.84964549,0.31207976){\color[rgb]{0,0,0}\makebox(0,0)[t]{\lineheight{1.29999995}\smash{\begin{tabular}[t]{c}Robot B\end{tabular}}}}%
    \put(0,0){\includegraphics[width=\unitlength,page=9]{figures/isac_1.pdf}}%
    \put(0.80575178,0.24210557){\color[rgb]{0,0,0}\makebox(0,0)[t]{\lineheight{1.29999995}\smash{\begin{tabular}[t]{c}Fire\end{tabular}}}}%
    \put(0,0){\includegraphics[width=\unitlength,page=10]{figures/isac_1.pdf}}%
    \put(0.58847974,0.26309269){\color[rgb]{0,0,0}\makebox(0,0)[t]{\lineheight{1.29999995}\smash{\begin{tabular}[t]{c}VaporISAC \end{tabular}}}}%
    \put(0,0){\includegraphics[width=\unitlength,page=11]{figures/isac_1.pdf}}%
    \put(0.54872769,0.05272071){\color[rgb]{0,0,0}\makebox(0,0)[t]{\lineheight{1.29999995}\smash{\begin{tabular}[t]{c}VaporISAC - passes thorugh cracks and gaps in debris \end{tabular}}}}%
    \put(0,0){\includegraphics[width=\unitlength,page=12]{figures/isac_1.pdf}}%
    \put(0.55915166,0.09766606){\color[rgb]{0,0,0}\makebox(0,0)[t]{\lineheight{1.29999995}\smash{\begin{tabular}[t]{c}VaporISAC - bypasses large boulders via fluid dynamics \end{tabular}}}}%
    \put(0,0){\includegraphics[width=\unitlength,page=13]{figures/isac_1.pdf}}%
    \put(0.60374223,0.1390144){\color[rgb]{0,0,0}\makebox(0,0)[t]{\lineheight{1.29999995}\smash{\begin{tabular}[t]{c}EM-ISAC (RF/LiDAR) - scattered by debris, absorbed by smoke \end{tabular}}}}%
    \put(0,0){\includegraphics[width=\unitlength,page=14]{figures/isac_1.pdf}}%
    \put(0.58945993,0.18168059){\color[rgb]{0,0,0}\makebox(0,0)[t]{\lineheight{1.29999995}\smash{\begin{tabular}[t]{c}EM-ISAC (RF/LiDAR) - blocked by tunnel roof for all robots \end{tabular}}}}%
    \put(0,0){\includegraphics[width=\unitlength,page=15]{figures/isac_1.pdf}}%
    \put(0.259059,0.85873783){\color[rgb]{0,0,0}\makebox(0,0)[t]{\lineheight{1.29999995}\smash{\begin{tabular}[t]{c}GNSS/GPS \end{tabular}}}}%
    \put(0,0){\includegraphics[width=\unitlength,page=16]{figures/isac_1.pdf}}%
    \put(0.60399693,0.85873783){\color[rgb]{0,0,0}\makebox(0,0)[t]{\lineheight{1.29999995}\smash{\begin{tabular}[t]{c}GNSS/GPS\end{tabular}}}}%
    \put(0,0){\includegraphics[width=\unitlength,page=17]{figures/isac_1.pdf}}%
    \put(0.95189902,0.8583299){\color[rgb]{0,0,0}\makebox(0,0)[t]{\lineheight{1.29999995}\smash{\begin{tabular}[t]{c}GNSS/GPS\end{tabular}}}}%
  \end{picture}%
\endgroup%

%% file: figures/isac_2.pdf_tex
\begingroup%
  \makeatletter%
  \providecommand\color[2][]{%
    \errmessage{(Inkscape) Color is used for the text in Inkscape, but the package 'color.sty' is not loaded}%
    \renewcommand\color[2][]{}%
  }%
  \providecommand\transparent[1]{%
    \errmessage{(Inkscape) Transparency is used (non-zero) for the text in Inkscape, but the package 'transparent.sty' is not loaded}%
    \renewcommand\transparent[1]{}%
  }%
  \providecommand\rotatebox[2]{#2}%
  \newcommand*\fsize{\dimexpr\f@size pt\relax}%
  \newcommand*\lineheight[1]{\tiny\fontsize{\fsize}{#1\fsize}\selectfont}%
  \ifx\svgwidth\undefined%
    \setlength{\unitlength}{495.16747662bp}%
    \ifx\svgscale\undefined%
      \relax%
    \else%
      \setlength{\unitlength}{\unitlength * \real{\svgscale}}%
    \fi%
  \else%
    \setlength{\unitlength}{\svgwidth}%
  \fi%
  \global\let\svgwidth\undefined%
  \global\let\svgscale\undefined%
  \makeatother%
  \begin{picture}(1,0.7020712)%
    \lineheight{1}%
    \setlength\tabcolsep{0pt}%
    \put(0,0){\includegraphics[width=\unitlength,page=1]{figures/isac_2.pdf}}%
    \put(0.13717653,0.06688287){\color[rgb]{0,0,0}\makebox(0,0)[t]{\lineheight{1.29999995}\smash{\begin{tabular}[t]{c}Time\end{tabular}}}}%
    \put(0,0){\includegraphics[width=\unitlength,page=2]{figures/isac_2.pdf}}%
    \put(0.13715273,0.24314996){\color[rgb]{0,0,0}\makebox(0,0)[t]{\lineheight{1.29999995}\smash{\begin{tabular}[t]{c}Robot Tx\end{tabular}}}}%
    \put(0,0){\includegraphics[width=\unitlength,page=3]{figures/isac_2.pdf}}%
    \put(0.05473081,0.15033103){\color[rgb]{0,0,0}\rotatebox{90}{\makebox(0,0)[t]{\lineheight{1.29999995}\smash{\begin{tabular}[t]{c}Ampltitude\end{tabular}}}}}%
    \put(0,0){\includegraphics[width=\unitlength,page=4]{figures/isac_2.pdf}}%
    \put(0.13476817,0.43560311){\color[rgb]{0,0,0}\makebox(0,0)[t]{\lineheight{1.29999995}\smash{\begin{tabular}[t]{c}Modulated Data Bits\end{tabular}}}}%
    \put(0,0){\includegraphics[width=\unitlength,page=5]{figures/isac_2.pdf}}%
    \put(0.13764311,0.5532322){\color[rgb]{0,0,0}\makebox(0,0)[t]{\lineheight{1.29999995}\smash{\begin{tabular}[t]{c}Inter-Robot Messages\end{tabular}}}}%
    \put(0,0){\includegraphics[width=\unitlength,page=6]{figures/isac_2.pdf}}%
    \put(0.49324927,0.586404){\color[rgb]{0,0,0}\makebox(0,0)[t]{\lineheight{1.29999995}\smash{\begin{tabular}[t]{c}Airflow\end{tabular}}}}%
    \put(0,0){\includegraphics[width=\unitlength,page=7]{figures/isac_2.pdf}}%
    \put(0.49586102,0.3244554){\color[rgb]{0,0,0}\makebox(0,0)[t]{\lineheight{1.29999995}\smash{\begin{tabular}[t]{c}Turbulence\end{tabular}}}}%
    \put(0,0){\includegraphics[width=\unitlength,page=8]{figures/isac_2.pdf}}%
    \put(0.13643855,0.66213594){\color[rgb]{0,0,0}\makebox(0,0)[t]{\lineheight{1.29999995}\smash{\begin{tabular}[t]{c}Transmitter\end{tabular}}}}%
    \put(0,0){\includegraphics[width=\unitlength,page=9]{figures/isac_2.pdf}}%
    \put(0.49389349,0.66213594){\color[rgb]{0,0,0}\makebox(0,0)[t]{\lineheight{1.29999995}\smash{\begin{tabular}[t]{c}Channel\end{tabular}}}}%
    \put(0,0){\includegraphics[width=\unitlength,page=10]{figures/isac_2.pdf}}%
    \put(0.86552319,0.65992409){\color[rgb]{0,0,0}\makebox(0,0)[t]{\lineheight{1.29999995}\smash{\begin{tabular}[t]{c}Receiver\end{tabular}}}}%
    \put(0,0){\includegraphics[width=\unitlength,page=11]{figures/isac_2.pdf}}%
    \put(0.79349908,0.58149498){\color[rgb]{0,0,0}\makebox(0,0)[t]{\lineheight{1.29999995}\smash{\begin{tabular}[t]{c}Ambient\end{tabular}}}}%
    \put(0.79399942,0.55367454){\color[rgb]{0,0,0}\makebox(0,0)[t]{\lineheight{1.29999995}\smash{\begin{tabular}[t]{c}State\end{tabular}}}}%
    \put(0,0){\includegraphics[width=\unitlength,page=12]{figures/isac_2.pdf}}%
    \put(0.85832802,0.24332846){\color[rgb]{0,0,0}\makebox(0,0)[t]{\lineheight{1.29999995}\smash{\begin{tabular}[t]{c}Robot Rx\end{tabular}}}}%
    \put(0,0){\includegraphics[width=\unitlength,page=13]{figures/isac_2.pdf}}%
    \put(0.49494288,0.45912668){\color[rgb]{0,0,0}\makebox(0,0)[t]{\lineheight{1.29999995}\smash{\begin{tabular}[t]{c}Smoke\end{tabular}}}}%
    \put(0,0){\includegraphics[width=\unitlength,page=14]{figures/isac_2.pdf}}%
    \put(0.86868852,0.4358173){\color[rgb]{0,0,0}\makebox(0,0)[t]{\lineheight{1.29999995}\smash{\begin{tabular}[t]{c}Joint Decoder\end{tabular}}}}%
    \put(0,0){\includegraphics[width=\unitlength,page=15]{figures/isac_2.pdf}}%
    \put(0.93129693,0.55430871){\color[rgb]{0,0,0}\makebox(0,0)[t]{\lineheight{1.29999995}\smash{\begin{tabular}[t]{c}Data Bits\end{tabular}}}}%
    \put(0,0){\includegraphics[width=\unitlength,page=16]{figures/isac_2.pdf}}%
    \put(0.51906348,0.06746446){\color[rgb]{0,0,0}\makebox(0,0)[t]{\lineheight{1.29999995}\smash{\begin{tabular}[t]{c}Time\end{tabular}}}}%
    \put(0,0){\includegraphics[width=\unitlength,page=17]{figures/isac_2.pdf}}%
    \put(0.39703957,0.15033103){\color[rgb]{0,0,0}\rotatebox{90}{\makebox(0,0)[t]{\lineheight{1.29999995}\smash{\begin{tabular}[t]{c}Ampltitude\end{tabular}}}}}%
    \put(0,0){\includegraphics[width=\unitlength,page=18]{figures/isac_2.pdf}}%
    \put(0.8886321,0.06746446){\color[rgb]{0,0,0}\makebox(0,0)[t]{\lineheight{1.29999995}\smash{\begin{tabular}[t]{c}Time\end{tabular}}}}%
    \put(0,0){\includegraphics[width=\unitlength,page=19]{figures/isac_2.pdf}}%
    \put(0.95591054,0.14534431){\color[rgb]{0,0,0}\makebox(0,0)[t]{\lineheight{1.29999995}\smash{\begin{tabular}[t]{c}$\theta$\end{tabular}}}}%
    \put(0,0){\includegraphics[width=\unitlength,page=20]{figures/isac_2.pdf}}%
    \put(0.76660963,0.15033103){\color[rgb]{0,0,0}\rotatebox{90}{\makebox(0,0)[t]{\lineheight{1.29999995}\smash{\begin{tabular}[t]{c}Ampltitude\end{tabular}}}}}%
    \put(0,0){\includegraphics[width=\unitlength,page=21]{figures/isac_2.pdf}}%
  \end{picture}%
\endgroup%